\documentstyle[preprint,aps,pre,psfig]{revtex}
\tightenlines
\newcommand{\be}{\begin{equation}}
\newcommand{\ee}{\end{equation}}
\newcommand{\ba}{\begin{eqnarray}}
\newcommand{\ea}{\end{eqnarray}}
\newcommand{\bsigma}{\mbox{\boldmath $\sigma$}}

\newcommand{\br}{{\bf r}}

\newcommand{\bv}{{\bf v}}
\newcommand{\eps}{\epsilon}

\newcommand{\bxi}{\mbox{\boldmath $\xi$}}
\begin{document}
\title{Velocity Distributions in Homogeneously Cooling
and Heated Granular Fluids}
\author{
T.P.C.~van Noije and M.H.~Ernst\\
{\it Instituut voor Theoretische Fysica, Universiteit Utrecht,
Postbus 80006,
3508 TA Utrecht, The Netherlands}}
\date{\today}
\maketitle
\begin{abstract}
We study the 
single particle velocity distribution
for a granular fluid of inelastic hard
spheres or disks, using the Enskog-Boltzmann equation,
both for the homogeneous cooling of a freely evolving system
and for the stationary state of a uniformly heated system,
and explicitly calculate the fourth cumulant of the distribution.
For the undriven case, our result agrees well with computer simulations
of Brey et al.\ \cite{brey}.
Corrections due to non-Gaussian behavior on cooling rate and  
stationary temperature 
are found to be small at all inelasticities.
The velocity distribution in the uniformly heated steady state
exhibits a high energy tail $\sim \exp(-A
c^{3/2})$, where $c$ is the velocity scaled by the thermal velocity and
$A\sim 1/\sqrt{\eps}$ with $\eps$ the inelasticity.
\end{abstract}
\narrowtext
\section{Introduction}
Most theories for rapid granular flows are based on the assumption that
the particle velocities are distributed according to a Gaussian or Maxwell
distribution.
Since granular particles collide inelastically, this assumption is not
an obvious one, however.
In fact, granular systems are typically far from equilibrium
systems in the sense that an external driving force is necessary to maintain a
stationary or periodic state.
Only in systems of nearly elastic particles states 
close to equilibrium are feasible.

Non-Gaussian behavior in rapid granular flows has been studied in
several contexts.
Taguchi and Hayakawa 
\cite{taguchi} observed power law behavior of the tails of the
velocity distribution in computer simulations of a bed of grains 
fluidized by vertical
vibration, and were able to explain their observations.
Esipov and P\"oschel \cite{esipov} have solved the 
Boltzmann equation
for inelastic hard spheres or disks for large velocities.
For the freely evolving system, they found a spatially
homogeneous distribution $f$ 
which for large velocities
decays as 
$f\sim \exp(-A v/v_0(t))$, 
where $v_0(t)$ is the time dependent
thermal velocity and $A\sim 1/\eps$ a constant, related to the
inelasticity $\eps=1-\alpha^2$, defined in terms of  
the coefficient of normal restitution $\alpha$.
An enhanced population for large energies was also found by Brey
et al.\ \cite{brey2}, who, based on a BGK model kinetic equation for
an undriven granular gas, found 
algebraically decaying tails with diverging velocity moments of
degree $\ge 2/\eps$. 
Sela and Goldhirsch \cite{sela} 
have numerically obtained a perturbative solution of the
Boltzmann equation for inelastic hard spheres 
to orders of ${\cal O}(\eps)$, ${\cal O}(\eps k)$,
${\cal O} (k^2)$.
The order ${\cal O}(\eps)$ estimates the deviation from 
Gaussian behavior of the
homogeneous 
solution and contributes to the rate of homogeneous
cooling.

Here we will present an alternative approach for solving the 
Enskog-Boltzmann equation for homogeneous single particle distributions.
We calculate explicitly the fourth moment for a freely evolving
system
of inelastic hard spheres or disks,
where the distribution function obeys a
scaling form, i.e. it is time dependent via the decaying temperature $T(t)$ 
only.
Our method follows Goldshtein and Shapiro \cite{goldshtein},
who calculated the fourth moment for a freely evolving gas of
inelastic hard spheres
($d=3$), but unfortunately made an error in their algebra.
From the fourth cumulant, we obtain the corrected cooling rate.

Next we consider a system of inelastic hard disks or spheres, where
at equal times a random velocity is added to the velocity of
each particle, referred to as uniformly heated system or as 
random acceleration model.
This idea of uniform heating allows for the existence of a homogeneous
stationary state and was first introduced for inelastic
particles on a line by Williams and MacKintosh\cite{williams}.
Recently, Peng and Ohta \cite{peng} and Puglisi at al.\
\cite{puglisi}
have performed
computer simulations of two-dimensional
and one-dimensional systems, respectively.
For the steady state we calculate the fourth cumulant of the velocity
distribution, 
as well as the corresponding stationary temperature.
Moreover we will present a calculation of the high energy tail of the
distribution function for
the uniformly heated system similar to the one presented in Ref.\
\cite{esipov} for an undriven granular gas.

We solve the equation for the single particle velocity 
distribution in the homogeneous state by expanding it in Sonine 
polynomials and deriving
equations for its moments.
As the calculation of the fourth moment is already quite
involved, we have not attempted to calculate higher moments.
Comparing our calculation with the one presented in Ref.\
\cite{sela}, we note that the results for the
${\cal O}(\eps)$ correction to the cooling rate are very
close.
Whereas the calculation in Ref.\ \cite{sela}
provides quantitative information on the
distribution function itself, ours provides quantitative
information on its moments, and shows only qualitatively similar
behavior for the distribution. 
The advantage of our method, however, is that it is nonperturbative in
the inelasticity, i.e.\ the moments can be obtained for all values of
the coefficient of restitution.

Our starting point is the nonlinear Enskog-Boltzmann equation \cite{ring}
for 
the single
particle distribution function 
$f(\br,\bv,t)$ in a dense system of inelastic hard spheres in $d$
dimensions.
In the absence of external forces, the {\em homogeneous} solution $f(\bv,t)$ 
of this equation obeys 
\ba
\partial_t f(\bv_1,t)&=&\chi \sigma^{d-1} \int {\rm d}{\bf v}_2 \int^\prime
{\rm d}\hat{\bsigma} ({\bf v}_{12}\cdot\hat{\bsigma})
\left\{\frac{1}{\alpha^2}f({\bf v}_1^{\ast\ast},t)f(
{\bf v}_2^{\ast\ast},t)-f({\bf v}_1,t)f({\bf v}_2,t)
\right\}\nonumber\\
&\equiv& \chi I(f,f).
\label{eq:boltz}
\ea
The prime on the $\hat{\bsigma}$ integration denotes the condition
${\bf
v}_{12}\cdot\hat{\bsigma}>0$, where $\hat{\bsigma}$ is a unit vector
along the line of centers of the colliding spheres at contact.
In {\em direct} collisions of inelastic hard spheres with a
coefficient of normal restitution $\alpha$,
the initial relative velocity ${\bf v}_{12}$ follows
the inelastic reflection law ${\bf
v}_{12}^\ast\cdot\hat{\bsigma}=-\alpha {\bf
v}_{12}\cdot\hat{\bsigma}$.
The gain term in (\ref{eq:boltz}) describes the {\em restituting}
collisions, i.e.\ the precollision velocities
($\bv_1^{\ast\ast},\bv_2^{\ast\ast}$) yield ($\bv_1,\bv_2$) as
postcollision ones with ${\bf
v}_{12}^{\ast\ast}\cdot\hat{\bsigma}=-(1/\alpha) {\bf
v}_{12}\cdot\hat{\bsigma}$.
Total momentum is conserved in a binary collision, and consequently
in direct collisions
\ba
{\bf v}^{\ast}_1&=&{\bf
v}_1-\textstyle{\frac{1}{2}}(1+\alpha)({\bf
v}_{12}\cdot\hat{\bsigma})\hat{\bsigma}\nonumber\\
{\bf v}^{\ast}_2&=&{\bf
v}_2+\textstyle{\frac{1}{2}}(1+\alpha)({\bf
v}_{12}\cdot\hat{\bsigma})\hat{\bsigma},
\label{eq:collr}
\ea
whereas ${\bf v}_i^{\ast\ast}(\alpha)={\bf v}_i^\ast(1/\alpha)$ in
restituting collisions.
The factor $1/\alpha^2$ in the gain term originates from the Jacobian
${\rm d}\bv_1^{\ast\ast} {\rm d}\bv_2^{\ast\ast}=(1/\alpha) {\rm
d}\bv_1{\rm d}\bv_2$ and from the length of the collision cylinder
$|{\bf
v}_{12}^{\ast\ast}\cdot\hat{\bsigma}|{\rm d}t=(1/\alpha) |{\bf
v}_{12}\cdot\hat{\bsigma}|{\rm d}t$.

Note that for the spatially homogeneous case, the only difference between 
the Enskog-Boltzmann equation for dense systems and the Boltzmann equation 
for dilute systems, is the presence of the factor 
$\chi(n)$, which is the pair correlation function at contact.
It accounts for the increased collision frequency in dense systems,
caused by excluded volume effects.

For later reference we will also quote the equation for the rate of change 
of the average
$\langle \psi \rangle = (1/n) \int {\rm d}\bv \psi(\bv) f(\bv,t)$,
where the density $n=\int {\rm d}\bv f(\bv,t)$.
From (\ref{eq:boltz}) it follows as
\be
\frac{{\rm d}\langle \psi \rangle}{{\rm d}t}=\frac{\chi \sigma^{d-1}}{2
n} \int{\rm d}\bv_1 {\rm d}\bv_2 \int^\prime {\rm d}\hat{\bsigma}
({\bf v}_{12}\cdot\hat{\bsigma}) f({\bf v}_1,t)f({\bf v}_2,t)
\Delta [\psi(\bv_1)+\psi(\bv_2)],
\label{eq:eqc}
\ee
where $\Delta \psi(\bv_i)=\psi(\bv_i^\ast)-\psi(\bv_i)$ is the $\psi$
change in a direct collision.

In the next section we study the solution of (\ref{eq:boltz}) 
for a freely evolving fluid.
In the subsequent section a uniformly heated system of inelastic particles
will be considered.
\section{Homogeneous Cooling State}
For the freely evolving granular fluid, 
Goldshtein and Shapiro \cite{goldshtein} have shown that Eq.\ 
(\ref{eq:boltz}) admits an isotropic scaling solution, 
describing the homogeneous
cooling state, 
with a single 
particle
distribution function depending on time only through the temperature 
$T(t)$ as  
\be
f({\bf v},t)=\frac{n}{v_0^d(t)}\tilde{f}\left(\frac{
v}{v_0(t)}\right),
\label{eq:scaling}
\ee
where the thermal velocity $v_0(t)$ is defined in terms of the
temperature by 
$T(t)=\textstyle{\frac{1}{2}}m v_0^2(t)$,
with
\be
\textstyle{\frac{1}{2}}d n T(t)=\int {\rm d}\bv \textstyle{\frac{1}{2}} m v^2 f(\bv,t),
\label{eq:Tdef}
\ee
and $m$ the particle mass.
Choosing $\psi=\textstyle{\frac{1}{2}}m v_1^2$ in Eq.\ (\ref{eq:eqc}) we
obtain for the rate of change of the temperature 
\be
\frac{{\rm d}T}{{\rm d}t}=-\frac{\mu_2}{d}m \chi n\sigma^{d-1} v_0^3 
\equiv-2\gamma
\omega_0 T.
\label{eq:Trate}
\ee
Here $\omega_0$ is the Enskog collision frequency for elastic hard
spheres, defined as the average
loss term in Eq.\ (\ref{eq:boltz}),
\be
\omega_0=\chi n \sigma^{d-1} \left\langle \int^\prime {\rm d}\hat{\bsigma}
({\bf v}_{12}\cdot\hat{\bsigma}) \right\rangle_0 =
\frac{\Omega_d}{\sqrt{2\pi}} \chi n \sigma^{d-1} v_0,
\label{eq:Efreq}
\ee
where $\langle \dots \rangle_0$ denotes an average over Maxwellian
velocity distributions for $\bv_1$ and $\bv_2$ at temperature $T=\textstyle{\frac{1}{2}}m v_0^2$ and
$\Omega_d=2\pi^{d/2}/\Gamma(d/2)$ is the surface area of a
$d$-dimensional unit sphere.
The second equality in (\ref{eq:Trate}) defines the {\em time independent} 
dimensionless cooling rate as
$\gamma\equiv(\sqrt{2\pi}/d \Omega_d)\mu_2$, where
\be
\mu_p\equiv -\int {\rm d}{\bf c}_1 c_1^p \widetilde{I}(\tilde{f},\tilde{f})
\label{eq:mu}
\ee
are the moments of the dimensionless collision integral
\be
\widetilde{I}(\tilde{f},\tilde{f})\equiv\int {\rm d} {\bf c}_2
\int^\prime
{\rm d} \hat{\bsigma} ({\bf c}_{12}\cdot\hat{\bsigma})
\left\{
\frac{1}{\alpha^2} \tilde{f}(c_1^{\ast\ast})
\tilde{f}(c_2^{\ast\ast})-\tilde{f}(c_1)
\tilde{f}(c_2)\right\},
\label{eq:collterm}
\ee
with ${\bf c}=\bv/v_0(t)$.
Using  
Eqs.\ (\ref{eq:scaling}) and 
(\ref{eq:Trate}), the scaling form $\tilde{f}(c)$ satisfies 
the integral equation
\be
\frac{\mu_2}{d} \left(d+c_1\frac{\rm d}
{{\rm d} 
c_1}\right) \tilde{f}(c_1)=
\widetilde{I}(\tilde{f},\tilde{f}).
\label{eq:inteq2}
\ee

In the limit of small dissipation, the solution of (\ref{eq:inteq2}) 
approaches a Maxwellian, i.e.\ $\tilde{f}(c)\approx
\phi(c) \equiv \pi^{-d/2} \exp(-c^2)$.
Therefore, a systematic approximation of the isotropic function $\tilde{f}(c)$
can be found by expanding it  
in a set of Sonine polynomials, i.e.\
\be
\tilde{f}(c)=\phi(c)\left\{1+\sum_{p=1}^\infty a_p S_p(c^2)\right\},
\label{eq:expansion}
\ee
which satisfy the orthogonality relations
\be
\int {\rm d}{\bf c} \phi(c) S_p(c^2) S_{p^\prime}(c^2) = \delta_{pp^\prime} 
{\cal N}_p,
\ee
where $\delta_{pp^\prime}$ is the Kronecker delta and ${\cal N}_p$ a 
normalization 
constant.
For general dimensionality $d$, the first few Sonine polynomials are
\ba
&S_0(x)=1\nonumber\\
&S_1(x)=-x+\frac{1}{2}d\nonumber\\
&S_2(x)=\frac{1}{2}x^2-\frac{1}{2}(d+2)x+\frac{1}{8}d(d+2).
\ea
The coefficients $a_p$ are polynomial moments of the scaling function:
\be
a_p = \frac{1}{{\cal N}_p} \int {\rm d}{\bf c} S_p(c^2) \tilde{f}(c) \equiv 
\frac{1}{{\cal N}_p} \langle S_p(c^2)\rangle.
\ee
In particular $a_1=(2/d)\langle S_1(c^2)\rangle=0$, because the temperature 
definition (\ref{eq:Tdef}) implies 
$\langle c^2\rangle=\textstyle{\frac{1}{2}}d$.
Moreover, $a_2$ is proportional to the fourth cumulant 
of the scaling form $\tilde{f}(c)$, i.e.\
\be
a_2=\frac{4}{d(d+2)}\left[\langle c^4\rangle-\textstyle{\frac{1}{4}}d(d+2)\right]=
\textstyle{\frac{4}{3}}\left[\langle c_x^4\rangle -3{\langle c_x^2\rangle}^2\right],
\label{eq:cumul}
\ee
where we have used the relation,
$\langle c_x^4\rangle=3\langle c^4\rangle/[d(d+2)]$, valid for any isotropic 
distribution $\tilde{f}(c)$.

To determine the coefficients $a_p$ we construct a set of equations 
for the moments
\be
\langle c^p\rangle \equiv\int {\rm d}{\bf c} c^p \tilde{f}(c),
\label{eq:mom}
\ee
by multiplying (\ref{eq:inteq2}) with $c_1^p\,\, (p=1,2,\dots)$ and 
integrating over ${\bf c}_1$.
For the moments $\mu_p$, defined in Eq.\ (\ref{eq:mu}), we obtain 
\ba
\mu_p&=&-\frac{\mu_2}{d}\int {\rm d}{\bf c} c^p \left(d +c\frac{\rm d}{{\rm d}c} \right) \tilde{f}(c)\nonumber\\
&=& \frac{\mu_2}{d} p \langle c^p\rangle,
\label{eq:mu2}     
\ea
where the second line has been obtained by partial integration.
For $p=2$ the above equation reduces to a trivial identity because of 
the definition of temperature.

\begin{figure}[h]
\centerline{\psfig{figure=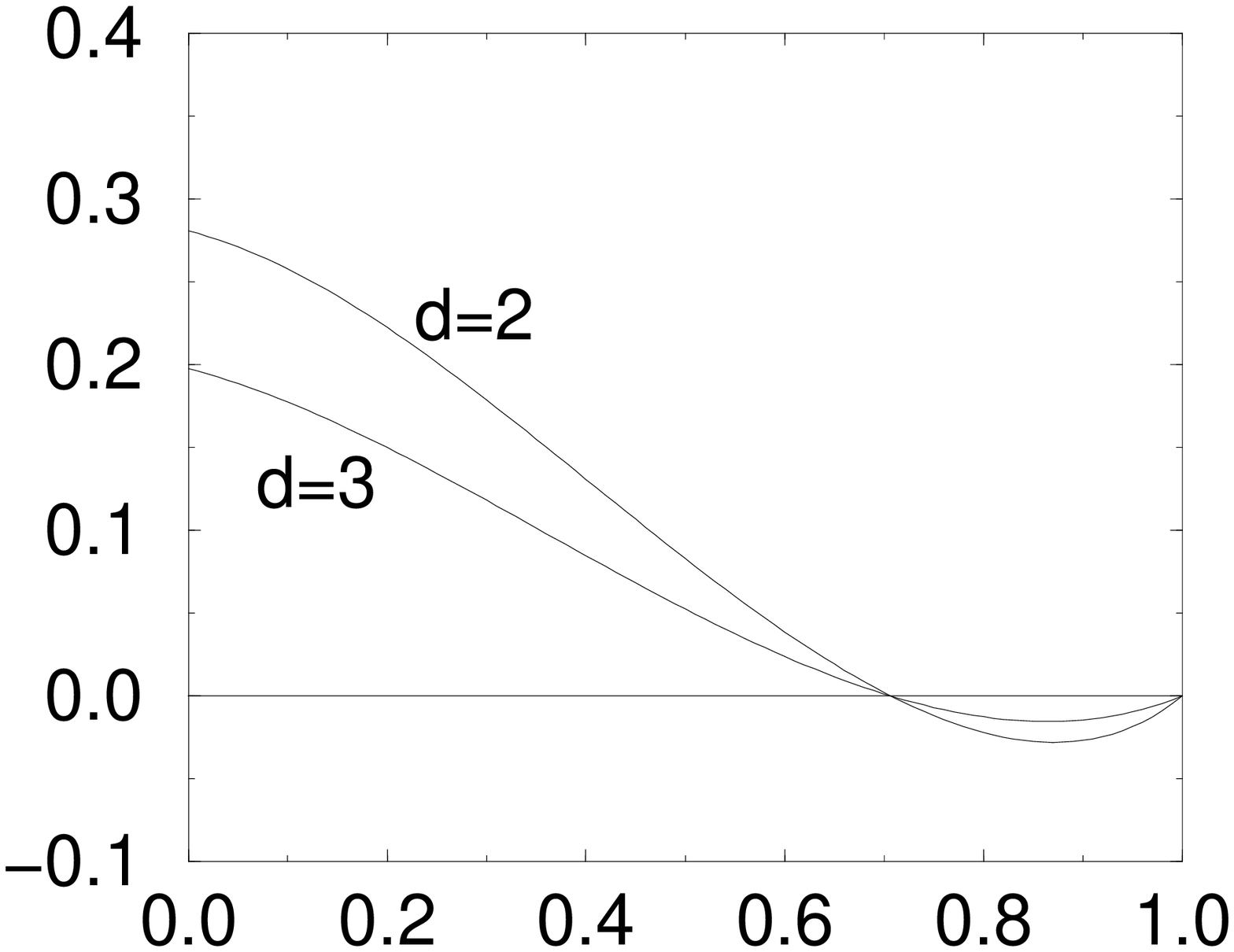,height=7cm}}
\caption{Fourth cumulant $a_2$ versus $\alpha$ for homogeneous cooling solution in a freely evolving fluid.}
\label{fig:a2}
\end{figure}

The quantities $\mu_2$, $\mu_p$ and $\langle c^p\rangle$ all depend on the unknown scaling function $\tilde{f}(c)$.
To calculate $a_2$ from (\ref{eq:mu2}) we set $p=4$, approximate the scaling form by $\tilde{f}(c)=\phi(c)\left\{1+a_2 S_2(c^2)\right\}$, and evaluate $\mu_2$, $\mu_4$ and $\langle c^4\rangle$.
The procedure is explained in more detail in the appendix and yields for general dimensionality $d$:
\be
a_2=\frac{16(1-\alpha)(1-2\alpha^2)}{9+24 d+8\alpha d-41 \alpha+30(1-\alpha)\alpha^2}.
\label{eq:a2}
\ee 
This result for $a_2$ is plotted in Fig.\ \ref{fig:a2} as a function
of $\alpha$.

In principle one can continue this approximation scheme by setting $\tilde{f}(c)=\phi(c)\left\{1+a_2 S_2(c^2)+ a_3 S_3(c^2)\right\}$, and then using (\ref{eq:mu2}) for $p=4$ and $p=6$ to obtain two coupled equations for $a_2$ and $a_3$, 
and solve the resulting equations to obtain better approximations for $a_2$ and $a_3$ than the previous ones, i.e.\ $a_2$ in (\ref{eq:a2}) and $a_3=0$.
As $a_2$ is already quite small, we do not calculate any higher coefficients $a_p\,\, (p\ge 3)$ in (\ref{eq:expansion}).

For the three-dimensional case Goldshtein and Shapiro \cite{goldshtein} have calculated the coefficient $a_2$ and find the result
\be
a_2^{\rm GS}=\frac{16(1-\alpha)(1-2\alpha^2)}{401-337\alpha+190 
(1-\alpha)\alpha^2}.
\ee
This result 
is only correct to linear order in 
$1-\alpha$ as the authors made an error in their algebraic
calculations\footnote{In the unpublished
appendices to their article, Eq.\ (E.10) should read $A_2=96+90
a_2$.}.
Their coefficient $|a_2^{\rm GS}|\lesssim 0.04$ for all $\alpha \in (0,1)$, 
whereas the correct coefficient obeys $|a_2|\lesssim 0.2$ for all $\alpha$.
However, the conclusion of Ref.\ \cite{goldshtein} that the homogeneous
scaling
form is well approximated by a Maxwellian remains valid for a large
range of coefficients of restitution (say $0.6\lesssim \alpha < 1$).
For these values we have $|a_2|\lesssim 0.04$ in three dimensions and
$|a_2|\lesssim
0.024$ in two dimensions.
Our result for $a_2$ has been quantitatively confirmed by the Direct
Simulation Monte Carlo results of Brey et al.\ \cite{brey}.
This will be discussed in section \ref{sec:sim}.

To obtain the time dependence of the temperature, it is convenient to
introduce the new time variable $\tau$ representing
 the average number of collisions
suffered per particle in a time $t$, and defined as
${\rm d}\tau=\omega_0(T(t)){\rm d}t$.
This yields
\be
\tau=\frac{1}{\gamma}\ln{(1+\gamma t/t_0)}.
\label{eq:tau}
\ee
Here $t_0=1/\omega_0(T_0)$ is the mean free time at the initial
temperature
$T(0)=T_0$.
Next we find from Eq.\
(\ref{eq:Trate})
\be
T(t)=T_0 \exp(-2 \gamma\tau)=\frac{T_0}{(1+\gamma t/t_0)^2}.
\label{eq:Tdecay}
\ee
In Eq.\ (\ref{eq:mu2app}) of the appendix, we derive for the cooling rate
$\gamma\equiv(\sqrt{2\pi}/d\Omega_d)\mu_2$:
\be
\gamma = \gamma_0\left\{1+\textstyle{\frac{3}{16}} a_2\right\},
\label{eq:gamma2}
\ee
where $\gamma_0=(1-\alpha^2)/2d$.
Sela and Goldhirsch \cite{sela} have performed a numerical perturbation 
expansion of
the Boltzmann equation to first order in $\eps=1-\alpha^2$ and found the result 
$\gamma=\gamma_0(1-0.0258\eps+{\cal O}(\eps^2))$, which is 
close to the result $\gamma=\gamma_0(1-3\eps/128+{\cal
O}(\eps^2))=\gamma_0(1-0.0234\eps+{\cal O}(\eps^2))$, obtained here. 
The method of appendix A also enables us to calculate the average
collision frequency $\omega=\omega[\tilde{f}]$ in the homogeneous scaling
state with the result
\be
\omega=\omega_0\left\{1-\textstyle{\frac{1}{16}} a_2\right\},
\label{eq:omega2}
\ee
where the Enskog frequency $\omega_0$ is defined in (\ref{eq:Efreq}).
Since the contribution from $a_2$ to $\gamma$ and $\omega$ are small
for all $\alpha$, (\ref{eq:gamma2}) and (\ref{eq:omega2}) are 
very well approximated by $\gamma_0$ and $\omega_0$, respectively.
\section{Uniformly Heated System}
To study this system we start from the  
stochastic equations of motion 
\be
\frac{{\rm d}{\bf v}_i}{{\rm d} t}= \frac{{\bf F}_i}{m} + \hat{\bxi}_i,
\label{eq:random}
\ee
where ${\bf F}_i$ is the force due to collisions and
$\hat{\bxi}_i$ is the random acceleration due to external forcing, which
is assumed to be Gaussian white noise and uncorrelated for different
particles, i.e.\
\be
\langle \hat{\xi}_{i\alpha}(t) \hat{\xi}_{j\beta}(t^\prime)\rangle =
\xi_0^2\delta_{ij}
\delta_{\alpha\beta}
\delta(t-t^\prime),
\ee
where $\xi_0^2$ is the strength of the correlation.
The validity of the above equations is based on the following
assumptions: (i) the system is thermodynamically large, so that
the condition $\sum_i \hat{\xi}_i(t)=0$, imposed in computer
simulations to guarantee
momentum conservation in finite systems, 
can be ignored; (ii) the time between random
kicks is small compared to the mean free time $t_0$, and therefore much smaller
than the characteristic cooling time $t_0/\gamma$ [see Eq.\ (\ref{eq:Tdecay})].

\begin{figure}
\centerline{\psfig{file=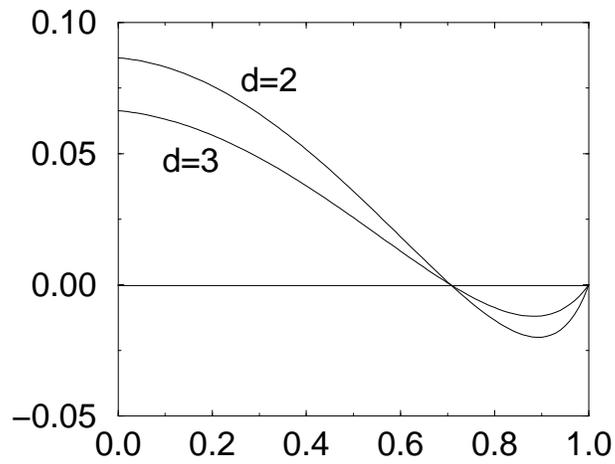,height=7cm}}
\caption{Fourth cumulant $a_2$ versus $\alpha$ for the 
stationary state of a uniformly heated system.}
\label{fig:a2h}
\end{figure}

The Enskog-Boltzmann equation for the single particle distribution function 
$f(\br,\bv,t)$ of a system heated in this way is corrected with
a Fokker-Planck diffusion term (see e.g.\ Ref.\ \cite{vankampen}), 
representing the change of the distribution function caused by the
small
random kicks,
and reads in the spatially homogeneous case:
\be
\partial_t f(\bv_1,t)=\chi I(f,f) +\frac{\xi_0^2}{2}
\left(\frac{\partial}{\partial{\bf v}_1}\right)^2 f(\bv_1,t).
\label{eq:fp}
\ee
The diffusion coefficient $\xi_0^2$ is proportional to the 
rate of energy input $\textstyle{\frac{d}{2}} \xi_0^2$ per
unit mass.
The equation for the temperature balance can be derived from Eq.\
(\ref{eq:fp}) in a similar fashion as in Eq.\ (\ref{eq:Trate}) for
the cooling granular fluid, and reads
\be
\frac{{\rm d} T}{{\rm d}t}=m \xi_0^2 -2\gamma\omega_0 T.
\label{eq:Tr}
\ee
We are looking for a stationary solution of (\ref{eq:fp}), 
where the heating exactly
balances the loss of energy due to collisions, and the temperature
becomes time independent.
Again it is convenient to introduce a scaled distribution function by
\be
f({\bf v})=\frac{n}{v_0^d}\tilde{f}\left(\frac{
v}{v_0}\right),
\ee
where now the thermal velocity $v_0$ is time independent.
Stationarity of $\tilde{f}$ then requires
\be
\widetilde{I}(\tilde{f},\tilde{f})+\frac{\xi_0^2}{2 v_0^3 \chi n
\sigma^{d-1}}
\left(\frac{\partial}{\partial{\bf c}_1}\right)^2  \tilde{f}(c_1)=0.
\label{eq:statf}
\ee
By multiplying this equation by $c_1^p$ and integrating over ${\bf c}_1$,
we obtain the following set of equations which couple the moments $\langle
c^{p-2} \rangle$ of the distribution to the moments $\mu_p$ of the
collision term, defined in Eq.\ (\ref{eq:mu}):
\be
\frac{\xi_0^2}{2 v_0^3 \chi n
\sigma^{d-1}} p (p+d-2) \langle
c^{p-2} \rangle = \mu_p.
\label{eq:rec}
\ee
For $p=2$ we recover the energy balance of Eq.\ (\ref{eq:Tr}),
yielding for the stationary 
value of the thermal velocity
in terms of $\mu_2$:
\be
v_0=\left(\frac{d \xi_0^2}{\mu_2 \chi n
\sigma^{d-1}}\right)^{1/3}.
\label{eq:stat}
\ee
Note that in order to obtain a finite temperature in the limit 
$\alpha\rightarrow
1$, the $\alpha$ limit should be taken together 
with the limit $\xi_0^2\rightarrow 0$.
The above expression is used to write Eq.\ (\ref{eq:rec}) in the form
\be
\frac{\mu_2}{2d} p(p+d-2) \langle
c^{p-2} \rangle = \mu_p.
\label{eq:rec1}
\ee
Since $a_1=0$ by definition of the temperature, i.e.\ $\langle
c^2\rangle=\textstyle{\frac{1}{2}}d$, the first correction to
Gaussian behavior is coming from $a_2$.
To calculate it, we take $p=4$ in Eq. (\ref{eq:rec1}), use expression 
(\ref{eq:mu4app}) for $\mu_4$, and solve for $a_2$ to finally obtain the
result
\be
a_2=\frac{16(1-\alpha)(1-2 \alpha^2)}{73
+56 d-24\alpha d -105 \alpha + 30(1-\alpha) \alpha^2}.
\label{eq:a2h}
\ee
This function is shown in Fig.\ \ref{fig:a2h} for the two- and
three-dimensional case.
Again we find only small corrections to a Maxwellian distribution
($a_2< 0.086$ in two dimensions and 0.067 in three).
Therefore to a good approximation, $\mu_2$ is given by its zeroth
order approximation and the stationary temperature is
found from Eqs. (\ref{eq:stat}) and (\ref{eq:mu2app}) as
\be
T_0=m \left(\frac{d\xi_0^2 \sqrt{\pi}}{(1-\alpha^2) \Omega_d \chi n
\sigma^{d-1}}\right)^{2/3}.
\label{eq:Tstat0}
\ee
\section{High Energy Tails}
In this section we will derive the asymptotic solution of the
Enskog-Boltzmann equation (\ref{eq:statf}) for high velocities in case the
granular fluid is uniformly heated.
Esipov and P\"oschel \cite{esipov} have given a similar derivation 
for a freely evolving gas and found a high energy tail 
$\tilde{f}(c)\sim \exp(-A c)$.
The derivation in both cases proceeds along similar lines.
If particle 1 is a fast particle ($c_1\gg 1$), 
the dominant contributions to the collision integral
are collisions where particle 2 is typically in the thermal range, 
so that 
${\bf c}_{12}$ in the collision integral
$\widetilde{I}(\tilde{f},\tilde{f})$ in (\ref{eq:collterm}) 
can be replaced by ${\bf c}_1$.
The gain term $\widetilde{I}_g$ of the collision
integral $\widetilde{I}$ can then be neglected with respect to the loss term
$\widetilde{I}_l$, as will be verified a posteriori
at the end of this section. 
The collision integral $\widetilde{I}(\tilde{f},\tilde{f})$ 
then reduces to $\widetilde{I}_l\approx -\beta_1 c_1 \tilde{f}(c_1)$, 
with
$\beta_1=\pi^{(d-1)/2}/\Gamma(\textstyle{\frac{1}{2}}(d+1))$ as given
in Eq.\ (\ref{eq:beta}) of the appendix,
and Eq.\ (\ref{eq:inteq2}) simplifies to
\be
\frac{\mu_2}{d} \left(d+c\frac{\rm d}
{{\rm d}
c}\right) \tilde{f}(c)=
-\beta_1 c \tilde{f}(c).
\ee
The first term on the left hand side can be neglected with respect to
the right hand side, and the large $c$ solution has the form
\be
\tilde{f}(c)\sim {\cal A}\exp(-\frac{\beta_1 d}{\mu_2} c),
\ee
where ${\cal A}$ is an undetermined integration constant. This
solution
corresponds to a tail which is
overpopulated when compared to
$\exp(-c^2)$.

To determine the high energy tail of $\tilde{f}(c)$ for the 
uniformly heated system, we proceed in a similar fashion and 
use (\ref{eq:stat}) to write
Eq.\ (\ref{eq:statf}) as
\be
\widetilde{I}(\tilde{f},\tilde{f})+\frac{\mu_2}{2 d}
\left(\frac{\partial}{\partial{\bf c}_1}\right)^2 
\tilde{f}(
c_1)=0.
\label{eq:statf1}
\ee
For large velocities $c_1$, the collision integral can again be
replaced by $- \beta_1 c_1 \tilde{f}(c_1)$, and
Eq.\ (\ref{eq:statf1}) reduces to
\be
- \beta_1 c \tilde{f}(c) + \frac{\mu_2}{2d} \left(\frac{{\rm
d}^2}{{\rm d} c^2} + \frac{d-1}{c} \frac{\rm
d}{{\rm d} c}\right) \tilde{f} (c)=0,
\ee
where we have used isotropy of the distribution function.
Inserting solutions of the form $\tilde{f}(c)\propto\exp(-A c^B)$, we
obtain the large $c$ solution with $B=\textstyle{\frac{3}{2}}$ and 
$A=\textstyle{\frac{2}{3}}\sqrt{\frac{2d
\beta_1}{\mu_2}}$, which is 
the only solution that vanishes for $c\rightarrow \infty$. 
Again we find an enhanced population for high energies.

To show that for $c_1\gg 1$ 
the gain term can be neglected with respect to the loss
term, we use the asymptotic collision dynamics
\ba
{\bf c}_1^{\ast\ast}&=& {\bf c}_1 -\textstyle{\frac{1}{2}}(1+\alpha^{-1})
({\bf
c}_1\cdot\hat{\bsigma})\hat{\bsigma}\nonumber\\
{\bf c}_2^{\ast\ast}&=& {\bf c}_2+\textstyle{\frac{1}{2}}(1+\alpha^{-1})
({\bf 
c}_1\cdot\hat{\bsigma})\hat{\bsigma},
\label{eq:adyn}
\ea
where we have replaced ${\bf c}_{12}$ by ${\bf c}_1$.
If $|{\bf
c}_1\cdot\hat{\bsigma}| \gg 1$, as is typically the case, 
${\bf c}_2$ in (\ref{eq:adyn}) can be neglected and we have
\ba
c_1^{\ast\ast}&=&c_1
\sqrt{1-\textstyle{\frac{1}{4}}(1+\alpha^{-1})(3-\alpha^{-1})
(\hat{\bf c}_1\cdot\hat{\bsigma})^2}\nonumber\\
c_2^{\ast\ast}&=&\textstyle{\frac{1}{2}}(1+\alpha^{-1})c_1 |\hat{\bf
c}_1\cdot\hat{\bsigma}| \gg 1,
\ea
where $\hat{\bf
c}_1$ is a unit vector. 
To demonstrate that $\tilde{f}(c)\sim \exp(-Ac^B)$ is a consistent
large $c$ solution, both in the freely evolving case with $B=1$ and
in the heated case with $B=\textstyle{\frac{3}{2}}$, we compare the
factor $\tilde{f}(c_1^{\ast\ast}) \tilde{f}(c_2^{\ast\ast})$ in
$\widetilde{I}_g$ with the factor $\tilde{f}(c_1)\tilde{f}(c_2)$ in
$\widetilde{I}_l$ for large $c$, i.e.\
\be
\frac{\tilde{f}(c_1^{\ast\ast})
\tilde{f}(c_2^{\ast\ast})}{\tilde{f}(c_1)\tilde{f}(c_2)}\sim \exp\left\{-A
[(c_1^{\ast\ast})^B+(c_2^{\ast\ast})^B-c_1^B]\right\}.
\label{eq:factor}
\ee
The exponent is proportional to $c_1^B$ and {\em strictly} negative
for $\alpha<1$ and $B<2$,
except for grazing collisions, where it vanishes.
This happens 
inside a
small $\theta$ interval $J$ of length ${\cal O}(1/c_1)$ near
$\theta=\pi/2$, where $|{\bf
c}_1\cdot\hat{\bsigma}|=c_1 \cos{\theta}\sim {\cal O}(1)$. 
Outside this interval the factor in (\ref{eq:factor}) vanishes
exponentially fast.
Inside the interval $J$ the factor in (\ref{eq:factor}) is ${\cal
O}(1)$. The contribution of this interval to the gain term can be
estimated as $\int_J {\rm d}\theta c_1
\cos{\theta}\tilde{f}(c_1)\simeq \tilde{f}(c_1)/c_1$, where $c_1
\cos{\theta}\sim {\cal O}(1)$.
Consequently, $\widetilde{I}_g/\widetilde{I}_l\sim 1/c_1^2$ for large
$c_1$.

In summary we have shown that $\tilde{f}(c)\sim \exp(-A c^B)$ is a
consistent large $c$ solution of the Boltzmann Eqs.\ (\ref{eq:inteq2})
and (\ref{eq:statf}) with $B=1$ for the freely evolving fluid and
$B=\textstyle{\frac{3}{2}}$ for the heated fluid.
\section{Comparison with Simulations}
\label{sec:sim}
In Refs.\ \cite{goldhirsch1,boston}, the undriven fluid of 
inelastic hard disks has been studied by
molecular dynamics simulations.
As long as the system 
is spatially homogeneous, measurements of the temperature decay 
confirm the validity of the homogeneous cooling law
(\ref{eq:Tdecay}) where the cooling rate $\gamma$ is given by its
zeroth order approximation $\gamma_0=\eps/2d$.
Also in the initial homogeneous state, 
the measured number of collisions $C$ among $N$ particles in a time 
$t$ is consistent with $\tau=2 C/N$ where $\tau$ is given by Eq.\
(\ref{eq:tau}), implying that the collision frequency $\omega$ is
very well approximated by its Enskog value $\omega_0$.

So far, molecular dynamics simulations have not been able to
obtain sufficient statistical accuracy to determine the
fourth moment or the high energy tail of the velocity distribution.
Such measurements are possible, however, by means of the Direct Simulation 
Monte
Carlo method for the Enskog-Boltzmann equation.
Using this method, Brey et al.\ \cite{brey} have
solved the nonlinear Boltzmann equation (\ref{eq:boltz}) for
homogeneously cooling inelastic hard spheres ($d=3$)
and measured the fourth and sixth moment of the
distribution $\tilde{f}(c)$.
Again the measured temperature decay shows no deviations of the
cooling rate $\gamma$ from its Gaussian value $\gamma_0$. 
Fig.\ 5 of Ref.\ \cite{brey} compares their simulation data for the fourth
cumulant $a_2$ with
(\ref{eq:a2}), first derived in \cite{noijeu}, and shows quantitative
agreement. 
In particular, we predict that the fourth cumulant vanishes for
$\alpha=1/\sqrt{2}$, which is very close to the value observed in the
simulations.
Also note that the simulation results disagree with the prediction of
Ref.\ \cite{goldshtein}.
Moreover, the approximation $\tilde{f}(c)=\phi(c)\{1+a_2 S_2(c^2)\}$
shows a good agreement with the simulation data for the functional
form of $\tilde{f}$ (see Figs.\ 7 and 8
of Ref.\ \cite{brey}).
This second Sonine approximation is qualitatively similar to the
form presented in Fig.\ 3 of Ref.\ \cite{sela}, calculated
numerically to order ${\cal O}(\eps)$.

It is also interesting to compare our theoretical predictions with
recent molecular dynamics results of Peng and Ohta \cite{peng} on the
{\em heated} granular fluid.
These authors have measured the temperature relaxation $T(t)$ in a
fluid of $N$ inelastic hard disks of mass $m=2$ at an area fraction
$\phi=\textstyle{\frac{1}{4}}\pi \sigma^2 N/L^2\simeq 0.16$ and
heating rate ${\xi_0^2}=(\delta V)^2/ 3 \tau_H\simeq 
1.67\times 10^{-4}$, where the randomly added velocity components
are sampled from a uniform distribution on the interval $(-\delta V,
\delta V)$ where $\delta V=10^{-3}$, measured in system length $L$
per unit time, and $\tau_H=2\times 10^{-3}$ is the period between
random kicks.
For the pair distribution of hard disks at contact, $\chi$, we use
the approximate form \cite{verlet} $\chi=(1-\textstyle{\frac{7}{16}}\phi)/(1-\phi)^2\simeq 1.32$.
The steady state temperature predicted by Eq.\ (\ref{eq:Tstat0})
then
becomes $T_0=1.15\times 10^{-3}$ for $\alpha=0.8$
The simulations yield $T_0^{\rm sim}\simeq 1.21\times 10^{-3}$ (see
Fig.\ 1 of Ref.\ \cite{peng}), in fair agreement with the Enskog
theory.

Moreover, Eq.\ (\ref{eq:Tstat0}) predicts that $T_0$ depends on the
heating rate $\xi_0^2=(\delta V)^2/3 \tau_H$ and
inelasticity $\eps=1-\alpha^2$, as 
\be
T_0=c_0\left(\frac{(\delta V)^2}{\tau_H(1-\alpha^2)}\right)^{2/3}\equiv
c_0 \zeta^\lambda.
\label{eq:para}
\ee
The measurements show an exponent $\lambda=0.65\pm 0.01$.
The theoretical prediction (\ref{eq:Tstat0}) gives $c_0\simeq
0.092$.
The simulation result of Ref.\ \cite{peng} $c_0^{\rm PO}\simeq 5.0
\times 10^{-3}$, corrected\footnote{The
parametrization (\ref{eq:para}) of the measurements in Ref.\
\cite{peng} with $c_0^{\rm PO}\simeq 5.0
\times 10^{-3}$ does not reproduce their data points at $\alpha=0.8$.
Moreover, the corresponding $\zeta$ value ($\zeta\simeq1.4\times
10^{-3}$) is far away from any data point included in their Fig.\ 6.}
with the conversion factor
$(L/\sigma)^{2/3}$, gives $c_0^{\rm sim}\simeq 5.0\times 10^{-3} \times
(70.9)^{2/3}\simeq 0.086$, again in fair agreement with the
theoretical prediction.

Eq.\ (\ref{eq:Tr}) also gives a prediction for the approach of $T(t)$
to $T_0$.
By observing that $\omega_0 \propto \sqrt{T}$ and {\em linearizing}
Eq.\ (\ref{eq:Tr})
around $T_0$, one obtains the solution $T(t)=T_0+T_1\exp(-t/\tau_0)$
with $\tau_0=2T_0/3 m \xi_0^2$.
For the parameters $\phi=0.16$ and $\alpha=0.8$ of Fig.\ 1 in Ref.\
\cite{peng} this yields $\tau_0=2.3$, and their simulations yield
$\tau_0^{\rm sim}=1/b^\prime=2.6$.

Next, we compare the collision frequency $\omega_0$ in (\ref{eq:Efreq})
from the Enskog theory for elastic hard disks with the collision
frequency $\omega^{\rm sim}$, measured in Ref.\ \cite{peng}.
If there are $C$ binary collisions among $N$ particles in a time $t$,
then the collision frequency is $2 C/N t$.
The simulation results at $\alpha=0.2$, 0.4, 0.6, 0.7 are respectively 
$\omega^{\rm sim}\simeq 2.93$, 1.67, 1.49, 1.49, and the Enskog predictions
for the same $\alpha$ values are $\omega_0\simeq 1.17$, 1.22, 1.33,
1.44.
The Enskog frequency $\omega_0\sim \sqrt{T_0}$ {\em decreases}
according to (\ref{eq:para}) with increasing $\eps=1-\alpha^2$,
whereas $\omega^{\rm sim}$ increases strongly with $\eps$.
The simulation results for $\omega^{\rm sim}$ suggest that the Enskog
theory gives reasonable predictions for $\alpha>0.6$. 
Similar conclusions have been obtained by Orza et al.\ \cite{boston}
for the homogeneous cooling state of a freely evolving fluid of 
inelastic hard disks.
The deviations at larger inelasticities are probably caused by
clustering and the onset of kinetic collapse, which strongly
increases the collision frequency.

Finally, we observe that the overpopulation in the high energy tail
$\sim \exp(-A
c^{3/2})$
of the steady state distribution function has also been observed in
the simulations of Ref.\ \cite{peng}, but their statistical accuracy
is too low to make any quantitative comparison.
\section{Conclusions and Outlook}
We have investigated non-Gaussian behavior
in 
granular fluids of smooth inelastic hard spheres or disks, both in the
absence of an 
external forcing and in a system uniformly heated by random accelerations.
In a freely evolving granular fluid, we find for all inelasticities
very small
corrections to the 
cooling rate and the collision
frequency due to non-Gaussian characteristics of the 
homogeneous cooling state.
As a consequence, such deviations have never been observed in 
computer simulations on homogeneous systems.
Our result for the fourth cumulant in the homogeneous cooling state
has been confirmed by the computer simulations of Brey et al.\
\cite{brey}.
These authors used the Direct Simulation Monte Carlo method to obtain
an accurate homogeneous solution of the nonlinear 
Enskog-Boltzmann equation.
This method might make it feasible to obtain
quantitative information on the high energy tail $\sim \exp(-A c)$ 
to test the theoretical predictions.
It certainly could be used to measure the fourth moment of the
homogeneous solution in a uniformly heated system, a quantity 
calculated in the present paper.
Again, in this case we predict very small corrections to the
stationary temperature and collision frequency due to
non-Gaussian properties of the homogeneous stationary state, 
which are possibly too small to measure in computer
simulations.
Moreover, it would be interesting to investigate the validity of our
prediction for the overpopulation of the high energy tail $\sim
\exp(-Ac^{3/2})$ with $A=\textstyle{\frac{2}{3}}\sqrt{\frac{2d
\beta_1}{\mu_2}}$.

Long range spatial correlations, measured in one- and
two-dimensional simulations of heated granular fluids 
\cite{williams,peng,puglisi},
are currently being analyzed using the ring kinetic
equations corresponding to kinetic equation (\ref{eq:fp}) with the
Fokker-Planck diffusion term, as well as by fluctuating hydrodynamics
with external noise.
The approach of adding {\em external} noise 
has some similarity with the Edwards-Wilkinson
model \cite{edwards} for the growth of a granular surface on which particles are
impinging at random.
As a consequence, long range spatial correlations in 
the velocity-velocity and 
density-density 
correlation functions are to be expected.

The clustering reported in the simulations of Ref.\ \cite{peng}, 
which causes  
an enhancement of the collision frequency 
with respect to the Enskog value for higher inelasticities
($\alpha\lesssim 0.6$), has not yet been explained in terms of
a linear or nonlinear stability analysis of the long wavelength
hydrodynamic modes of the system.
\\

T.v.N. acknowledges support of the
foundation `Fundamenteel Onderzoek der Materie (FOM)', which is
financially supported by the Dutch National Science Foundation
(NWO).
\section{Appendix A}
\setcounter{equation}{0}
\renewcommand{\theequation}{A.\arabic{equation}}
\label{sec:appendix}
In this appendix we calculate the quantities $\mu_2, \mu_4$ and $\langle c^4\rangle$, which are required in (\ref{eq:mu2}) to calculate the coefficient $a_2$ in (\ref{eq:expansion}) by setting $\tilde{f}(c)=\phi(c)\left\{1+a_2 S_2(c^2)\right\}$, where $\phi(c)$ is the Maxwellian.
In fact, the moment $\langle c^4\rangle$ in (\ref{eq:mom}) requires only moments of the Gaussian distribution.
A straightforward calculation gives
\ba
\langle c^4\rangle&=&\int {\rm d}{\bf c} c^4 \phi(c)\left\{1+a_2 S_2(c^2)\right\}\nonumber\\
&=&\textstyle{\frac{1}{4}} d(d+2) \left\{1+ a_2\right\}.
\label{eq:c4app}
\ea
Next we consider the moments $\mu_p\,\, (p=2, 4)$ in (\ref{eq:mu}).
With the help of Eq.\ (\ref{eq:eqc}) it can be transformed into
\ba
\mu_p &=&-\textstyle{\frac{1}{2}}\int {\rm d}{\bf c}_1\int {\rm d}{\bf c}_2 \int ^\prime {\rm d} \hat{\bsigma} ({\bf c}_{12}\cdot \hat{\bsigma}) \phi(c_1)\phi(c_2)\times\nonumber\\
&&\left\{1+a_2\left[S_2(c_1^2)+S_2(c_2^2)\right] + {\cal O}(a_2^2)\right\} \Delta(c_1^p+c_2^p),
\label{eq:muapp}
\ea
where the operator $\Delta$ is defined below Eq.\ (\ref{eq:eqc}).
In the following, terms of ${\cal O}(a_2^2)$ will be neglected.
For $\alpha\gtrsim 0.6$ this is a safe approximation as can be checked
from the results Eqs.\ (\ref{eq:a2}) and (\ref{eq:a2h}) for $a_2$.

To evaluate (\ref{eq:muapp}) we introduce center of mass and relative velocites by ${\bf c}_1={\bf C} + \textstyle{\frac{1}{2}}{\bf c}_{12}$ and ${\bf c}_2={\bf C}-\textstyle{\frac{1}{2}}{\bf c}_{12}$.
Moreover, we need the angular integral
\ba
\beta_n&\equiv& \int^\prime {\rm d}\hat{\bsigma} (\hat{\bf c}_{12}\cdot \hat{\bsigma})^n=\textstyle{\frac{1}{2}}\Omega_d \frac{\int ^\prime {\rm d}\hat{\bsigma}(\cos{\theta})^n}{\int^\prime {\rm d}\hat{\bsigma}}\nonumber\\
&=&\textstyle{\frac{1}{2}}\Omega_d \frac{\int_0^{\pi/2} {\rm d}\theta (\sin{\theta})^{d-2} (\cos{\theta})^n}{\int_0^{\pi/2} {\rm d}\theta (\sin{\theta})^{d-2}} =
\pi^{\frac{d-1}{2}} \frac{\Gamma(\frac{n+1}{2})}{\Gamma(\frac{n+d}{2})},
\label{eq:beta}
\ea
where $\hat{\bf c}_{12}={\bf c}_{12}/|{\bf c}_{12}|$ is a unit vector.
Using the relations between Gaussian moments, it is straightforward to derive the relations:
\ba
{\langle {\bf c}_{12}\cdot\hat{\bsigma} C^2 \Delta C^n c_{12}^m\rangle}_0 &=&
\textstyle{\frac{1}{4}}(d+n){\langle {\bf c}_{12}\cdot\hat{\bsigma}\Delta C^n c_{12}^m \rangle}_0\nonumber\\
{\langle {\bf c}_{12}\cdot\hat{\bsigma} C^4 \Delta C^n c_{12}^m \rangle}_0 &=&
\textstyle{\frac{1}{16}}(d+n)(d+n+2) {\langle {\bf c}_{12}\cdot\hat{\bsigma} \Delta C^n c_{12}^m\rangle}_0,
\label{eq:helpf}
\ea
where
\be
\langle \psi({\bf c}_{12},{\bf C}) \rangle_0 \equiv \int {\rm d} {\bf c}_{12}\frac{1}
{(2\pi)^{d/2}} \exp(-\textstyle{\frac{1}{2}}c_{12}^2)\int {\rm d}{\bf C}{\left(\frac{2}{\pi}\right)}^{d/2} \exp(-2 C^2)  \psi({\bf c}_{12},{\bf C})
\ee
denotes a Gaussian average over ${\bf c}_{12}$ and ${\bf C}$. 
The above formulas are very helpful in evaluating the moments $\mu_p$ in (\ref{eq:muapp}).
With the help of (\ref{eq:beta}) and (\ref{eq:helpf}) one finds 
\ba
\mu_2&=&\textstyle{\frac{1}{4}} (1-\alpha^2) \beta_3 \langle c_{12}^3\rangle_0\left\{1+\textstyle{\frac{3}{16}} a_2\right\}\nonumber\\
&=&\textstyle{\frac{1}{2}} (1-\alpha^2) \frac{\Omega_d}{\sqrt{2\pi}} \left\{1+\textstyle{\frac{3}{16}} a_2\right\}.
\label{eq:mu2app}
\ea

To calculate $\mu_4$ we need the quantity
\ba
\Delta(c_1^4+c_2^4)&=&2(1+\alpha)^2({\bf c}_{12}\cdot \hat{\bsigma})^2({\bf
C}\cdot \hat{\bsigma})^2+\textstyle{\frac{1}{8}}(\alpha^2-1)^2 ({\bf
c}_{12}\cdot \hat{\bsigma})^4\nonumber\\
&&+(\alpha^2-1)({\bf c}_{12}\cdot
\hat{\bsigma})^2C^2+\textstyle{\frac{1}{4}}(\alpha^2-1)({\bf c}_{12}\cdot
\hat{\bsigma})^2c_{12}^2\nonumber\\
&&-4(1+\alpha)({\bf c}_{12}\cdot
\hat{\bsigma})({\bf C}\cdot \hat{\bsigma})({\bf C}\cdot{\bf c}_{12}).
\ea
One finds after long and tedious calculations
\be 
\mu_4= \beta_3 \langle c_{12}^3\rangle_0\left\{T_1+a_2 T_2\right\},
\label{eq:mu4app}
\ee
with 
\ba
T_1&=&\textstyle{\frac{1}{4}} (1-\alpha^2)(d+\textstyle{\frac{3}{2}}+\alpha^2)\nonumber\\
T_2&=&\textstyle{\frac{3}{128}}(1-\alpha^2)(10d+39+10\alpha^2)+\textstyle{\frac{1}{4}}(1+\alpha)(d-1).
\ea
For the homogeneous cooling solution, inserting the results (\ref{eq:c4app}), (\ref{eq:mu2app}) and (\ref{eq:mu4app}) into (\ref{eq:mu2}) for $p=4$
yields a closed equation for $a_2$.
Neglecting again small contributions ${\cal O}(a_2^2)$, we solve for $a_2$, and the result in (\ref{eq:a2}) is recovered.
Eq.\ (\ref{eq:a2h}) corresponding to uniform heating is found by inserting
(\ref{eq:mu2app}) and (\ref{eq:mu4app}) into (\ref{eq:rec1}) for $p=4$.

\end{document}